\documentstyle[aps,preprint]{revtex}
\begin{document}
\preprint{U. of MD PP\#95--053}
\title{Exchange Current Corrections to Neutrino--Nucleus
Scattering\footnote{To appear in Physical Review Letters}}
\author{Y.~Umino\footnote{Present Address: Department of Physics,
University of Maryland, College Park, MD 20742-4111, U.S.A.},
J.M.~Udias and P.J.~Mulders\footnote{Also at Physics
Department, Free University, NL-1081 HV Amsterdam, The Netherlands.}}
\address{National Institute for Nuclear and High--Energy Physics,
Section K (NIKHEF--K), \\
Postbus 41882, NL--1009 DB Amsterdam, The Netherlands\\}
\date{\today}
\maketitle
\begin{abstract}
Relativistic exchange current corrections to
neutrino--nucleus cross sections are presented assuming non--vanishing strange
quark form factors for the constituent nucleons.
For charged current processes the exchange current corrections can lower
the impulse approximation results by 10\% while these corrections are
found to be sensitive to both the nuclear density and the strange quark
axial form factor of the nucleon for neutral current processes.
Implications on the LSND experiment to determine this form factor are
discussed.
\end{abstract}
It is well known that meson exchange currents play an essential role in
realistic descriptions of electroweak interactions in nuclei \cite{meson}.
For example in the electromagnetic sector, exchange current corrections
required
by gauge invariance are found to be important in explaining
the renormalization of orbital $g$--factors \cite{fuj71}, the threshold
radiative
neutron capture rates \cite{ris72}, or its inverse, the deuteron
photo-disintegration cross section
\cite{hoc73} and the transverse $(e,e')$ response functions in the dip
region \cite{ord81}. In addition, shell model studies of first forbidden
$\beta$-decay
rates covering a wide range of nuclei \cite{war} indicate
a substantial exchange current contribution to the renormalization of weak
axial charge in medium,
as predicted by Kubodera, Delorme and Rho in 1978 \cite{kub78}.
Thus, empirical evidences abound suggesting that both electromagnetic and weak
axial currents are
subject to renormalizations in nuclei due to exchange currents. It is therefore
interesting to examine the
effects of exchange currents, if any, in neutrino--nucleus scattering where
both vector and axial
currents are involved simultaneously.

Another reason to investigate exchange current corrections to neutrino--nucleus
scattering is that it has been receiving increasing attention as a means to
determine
the strangeness matrix elements of the nucleon [8--14].
The measurement of polarized structure function $g_1$ and the extraction of the
sum rule indicated
the possibility of a rather large strange quark axial matrix element for the
proton \cite{ell94}, and
has inspired numerous works attempting to understand the role of hidden flavor
in nucleons.
However, the situation regarding the
strangeness degrees of freedom in the nucleon is far from clear and
it is hoped that neutrino-nucleus interactions might be able to
shed a new light into this problem \cite{mus94}.
In order to extract strangeness matrix elements for the nucleon from
neutrino--nucleus scattering it is necessary to reliably calculate the
cross sections assuming finite strange quark form factors \cite{min84}.
The kinematics of neutrino--nucleus interactions involved in determining
the strange content of the nucleon ranges
from low-energy inelastic scattering \cite{suz90} to the quasi-elastic region
\cite{gar92}.
Experience from electron scattering suggests that exchange current
corrections to cross sections in this kinematic range might be important.

In this letter two-body exchange current corrections to the impulse
approximation
in low and intermediate energy neutrino--nucleus scattering are presented using
the
generalization of a method developed by Chemtob and Rho \cite{che71}.
As shown below this method is powerful enough to estimate
exchange current corrections to both neutral and charged current processes
simultaneously
assuming finite strange quark form factors of the nucleon.
In addition, the formalism involved in this approach is model independent in
the sense that
no nucleon--nucleon interaction need to be specified.
As examples, relativistic two-body exchange current corrections to
neutral and charged current neutrino--nucleus cross sections
are evaluated for several nuclear densities assuming nuclear matter and using
the
kinematics of the ongoing LSND  experiment to measure the strange axial form
factor of
the nucleon \cite{gar92}.

It is convenient to write the neutral and charged currents of a {\it free}
nucleon,
$J^{Z^0}_{\mu}$ and $J^{W^{\pm}}_{\mu}$,
in terms of $SU(3)$ vector, $V^a_{\mu}$, and axial, $A^a_{\mu}$, currents
where $a=0$ for singlet and $a=1 \rightarrow 8$ for octet currents,
respectively.
\begin{eqnarray}
J^{Z^0}_{\mu} & = & V^3_{\mu} - A^3_{\mu} - 2\, \sin^2\theta_W \left( V^3_{\mu}
+
\frac{1}{\sqrt{3}} V^8_{\mu} \right) - \frac{1}{2} \left( V^0_{\mu} -
\frac{2}{\sqrt{3}}V^8_{\mu} \right)
+ \frac{1}{2} \left( A^0_{\mu} - \frac{2}{\sqrt{3}}A^8_{\mu} \right)
\label{eq:JZERO} \\
J^{W^{\pm}}_{\mu}  & = & \biggl[ \Bigl( V^1_{\mu} \pm iV^2_{\mu} \Bigr)
- \Bigl( A^1_{\mu} \pm iA^2_{\mu} \Bigr) \biggr] \cos\theta_C
+ \biggl[ \Bigl( V^4_{\mu} \pm iV^5_{\mu} \Bigr)
- \Bigl( A^4_{\mu} \pm iA^5_{\mu} \Bigr) \Biggr] \sin\theta_C
\label{eq:JPM}
\end{eqnarray}
In these definitions of weak neutral and charged currents, $\theta_W$ and
$\theta_C$ are the Weinberg and Cabbibo angles, respectively, and
the last two terms in Eq.~(\ref{eq:JZERO})  are usually referred to as the
strange quark vector and axial currents of the nucleon.
Small QED, QCD and heavy quark corrections to $J^{Z^0}_{\mu}$  \cite{kap88} as
well as
contributions from the charmed quarks to $J^{W^{\pm}}_{\mu}$ are ignored.
Thus, the problem addressed in this letter is to estimate
exchange current corrections to the above currents when the nucleon is
immersed in nuclear medium.  Other medium effects, such as density dependent
off--shell
form factors and effective nucleon and meson masses, are not considered in this
work in order
to clearly isolate possible exchange current effects in many body systems.

The starting point is to assume the chiral--filtering
conjecture \cite{kub78} which states that the dominant exchange current
contribution in nuclei at low and intermediate energies comes
from the exchange of a single pion whose production amplitude is evaluated in
the soft--pion limit.
As mentioned, there exists a well--known method by Chemtob and Rho \cite{che71}
to construct soft--pion exchange current operators which exploits
soft--pion theorems and current algebra techniques pioneered by Adler
\cite{adl68}. There are several advantages of using this approach in solving
the problem
at hand. First, the method of Chemtob and Rho has been applied in the past
to various low and intermediate energy phenomena and proved
to be a reliable technique for estimating exchange current corrections
\cite{meson}.
For example in their analysis of first forbidden $\beta$--decay
transitions, Warburton {\it et al.} find that exchange current corrections
to the axial charge may reliably be estimated in the the soft--pion exchange
dominance
approximation \cite{war}.
Furthermore, most of the discrepancy between the measured deuteron
photo--disintegration
rates involving small energy and {\it large} momentum transfers
\cite{saclay} and its impulse approximation prediction can be explained by
soft--pion
exchange corrections \cite{hoc73}. This came as a surprise since the
soft--pion dominance approximation was thought to be applicable only to
processes involving small momentum transfers as in first forbidden
$\beta$--decay
transitions \cite{rho81}.

A justification for the success of the soft--pion exchange dominance for
finite momentum transfers was proposed by Rho who based his arguments on
Weinberg's
derivation of nuclear forces from chiral Lagrangians \cite{rho91}. Using chiral
power counting
Rho has argued that to the leading order, {\it i.e.} at the
tree level, the short range part of two--body meson exchange current
corresponding to a nuclear force predicted by a given chiral Lagrangian
is considerably suppressed. Thus the dominant contribution to two--body
currents comes from the long ranged part represented by the soft--pion
exchange.
Consequently, Park, Towner and Kubodera \cite{par94} have calculated
corrections to the axial charge exchange current operators beyond the
the soft--pion dominance approximation using heavy--fermion chiral
perturbation theory. They find that loop corrections to the soft--pion exchange
current
operators is of the order of 10\%, and argued that their results are
consistent with the claims of Warburton {\it et al.} and support
the chiral--filtering conjecture. Thus, the soft--pion
technique of constructing exchange current operators seems to be a
plausible approximation in both electromagnetic and weak axial sectors, and it
is natural to apply this method to neutrino--nucleus scattering where
both vector and axial currents are involved. Furthermore, there are
theoretical arguments suggesting that corrections to the soft--pion
dominance approximation are small, even for finite momentum transfers
extending to the quasi--elastic region \cite{rho81,rho91}.

Finally, the main advantage of using the method of Chemtob and Rho is
the equal treatment of all the vector and axial currents entering in neutral
and
charged currents as described below. The quantity of interest in this method is
the amplitude for pseudoscalar meson production off a nucleon
by an external current, denoted by $\langle N(p') \phi^b (q)| J_{\mu}^a(k) |
N(p) \rangle$.
Here $k$ and $q$ are the four momenta of the probing current $J_{\mu}^a$, and
the
produced meson $\phi^b$, respectively, while $a$ and $b$ are $SU(3)$ indices
($a = 0 \rightarrow 8$ while $b = 1, 2$ or $3$ for pion production).
In the present case $J_{\mu}^a$ may be any one of the $SU(3)$ vector or axial
currents appearing in Eqs.~(\ref{eq:JZERO}) and (\ref{eq:JPM}),
and the meson production amplitude is evaluated in the soft--meson limit
of $q \rightarrow 0$.
This soft--meson production amplitude, derived by Adler \cite{adl68} and used
by
Chemtob and Rho \cite{che71}, may be written in the generalized form as
\begin{eqnarray}
\lim_{q\rightarrow 0} \langle N(p') \phi^b (q)| J_{\mu}^a(k) | N(p) \rangle
& = &
\frac{i}{F_{\phi}} \int d^4 \! x \lim_{q\rightarrow 0} (-i q^{\nu})
\langle N(p')| T \left( A_{\nu}^b(x) J_{\mu}^a(0) \right) |N(p) \rangle
\nonumber \\
&    &
\;\;\;\;  - \frac{i}{F_{\phi}}
\langle N(p')| \left[ Q^b_5(x), J_{\mu}^a(0) \right]_{x_0=0} | N(p) \rangle
\label{eq:LSZ}
\end{eqnarray}
Here $Q^a_5(x) \equiv \int d^3\! x\, A_0^a(x)$ is the axial charge and
$F_{\phi}$ is the
decay constant for the pseudoscalar meson $\phi$.
As shown in \cite{adl65}, the only contributions to the first term in the
soft--meson limit come from pole terms where the matrix element
$\langle N(p')| T \left( A_{\nu}^b(x) J_{\mu}^a(0) \right) |N(p) \rangle$
behaves like $1/q_{\mu}$. The second term may be simplified by using the
well--known $SU(3) \otimes SU(3)$ current algebra
\begin{equation}
\left[ Q^a_5(x), V_{\mu}^b(0) \right]_{x_0=0} =  i f_{abc}
A_{\mu}^c(0) \;\;\;\;
\left[ Q^a_5(x), A_{\mu}^b(0) \right]_{x_0=0} =  i f_{abc} V_{\mu}^c(0)
\end{equation}
and has no contributions from singlet currents unlike in the first term
where both $SU(3)$ singlet and octet currents can contribute.
Since $J_{\mu}^a$ can be any of the $SU(3)$ vector or axial currents,
Eq.~(\ref{eq:LSZ}) may be applied to {\em all} the components of weak neutral
and charged currents
in Eqs.~(\ref{eq:JZERO}) and (\ref{eq:JPM}) simultaneously.
Thus, it is the use of current algebra in Eq.~(\ref{eq:LSZ}), which
rotates around the vector and axial octet currents, that makes this method
particularly suitable to estimate exchange current corrections in neutrino
scattering at low
and intermediate energies assuming finite strange quark form factors.

Two--body operators for neutral and charged--current induced exchange
currents may be constructed in a straightforward manner following
\cite{che71}. The non--relativistic limit of these operators have
previously been used to
calculate exchange current corrections to neutrino--deuteron scattering
in the original $SU(2)$ version supplemented by finite $q$ corrections
\cite{kubodera}.
In the present application, the operators are fully relativistic and the method
is
generalized to $SU(3)$ to accomodate finite strange quark form factors.
The conservation of the vector current has been checked both analytically
and numerically using the prescription outlined in \cite{adl68}.
Figure~1a shows differential cross sections for the neutral current
reaction $^{12}C(\nu, \nu' p)$ plotted against the kinetic energy of the
ejected nucleon $T_F$. The calculation was performed using the relativistic
Fermi Gas model formalism as in Horowitz {\it et al.} with zero binding
energy. Furthermore, to simulate
the LSND experiment, the nucleons are assumed to be ejected
quasi-elastically from the target nuclei by neutrinos with a beam energy of
200 MeV, and only $1p1h$ final states are considered when taking matrix
elements of two--body operators since the phase space for $2p2h$
final states should be highly suppressed for the LSND kinematics \cite{gar92}.

The differential cross sections are parameterized by the strange quark
magnetic,
$F_2^s \equiv F_2^s(Q^2=0)$, and axial, $G_A^s \equiv G_A^s(Q^2=0)$, form
factors of the nucleon. The $Q^2 \equiv -k^2$
dependence for $F_2^s$ and $G_A^s$ are assumed to be \cite{gar92}
\begin{equation}
F_2^s(Q^2)  \equiv  F_2^0(Q^2) - \frac{2}{\sqrt{3}}F_2^8(Q^2) \;\;\;\;
G_A^s(Q^2)  \equiv  G_A^0(Q^2) - \frac{2}{\sqrt{3}}G_A^8(Q^2)
\end{equation}
where
\begin{equation}
F_2^{0,8}(Q^2) \equiv
\frac{F_2^{0,8}(0)}{(1+\frac{Q^2}{4M_N^2})(1+\frac{Q^2}{M_V^2})^2} \;\;\;\;
G_A^{0,8}(Q^2)  \equiv
\frac{G_A^{0,8}(0)}{(1+\frac{Q^2}{M_A^2})^2}
\end{equation}
In these definitions $M_N$ is the nucleon mass and the vector and axial
masses are set to $M_V$ = 840 MeV and $M_A$ = 1030 MeV, respectively.
The octet form factors at $Q^2=0$ are known quantities given by $F_2^8(0)
\equiv \sqrt{3}/2(\kappa_p + \kappa_n)$ and $G_A^8(0) \equiv
\sqrt{3}/6(3F -D)$, while the unknown singlet form factors $F_2^0(0)$ and
$G_A^0(0)$ determine the strange quark form factors $F_2^s$ and $G_A^s$
which are assumed to be $F_2^s = -0.21$
and $G_A^s = -0.19$ for this work. The results obtained by assuming no strange
form factors
$(F_2^s = G_A^s = 0)$ are qualitatively similar but there is an overall
20\% reduction in the differential cross sections. The exchange current
corrections are found to be sensitive to the Fermi momentum  $k_F$ of the
model. For $k_F \approx$ 200 MeV there are cancellations between the
vector and axial contributions to the exchange current correction leading
to little change from the impulse approximation results. However, for
$k_F \approx$ 300 MeV and beyond there are considerable corrections to the
impulse approximation from exchange currents as shown in the figure.

Figure~1b shows the cross sections for the inclusive charged current
process $^{12}C(\nu_{\mu^-}, \mu^-)X$ for several nuclear densities
obtained by folding the LSND neutrino energy distribution \cite{alb94}.
Here the effect of exchange current corrections varies from 5 to 10\% as
the Fermi momentum is increased from 200 to 300 MeV. For $k_F$ = 225 MeV,
which is the usual value used for $^{12}C$, the total cross--section is
reduced from 24 to 22.7 $(\times 10^{-40} {\rm cm}^2)$. This reduction is
not enough to explain the recently measured value reported by the LSND
collaboration of $(8.3 \pm 0.7\; {\rm stat.} \pm 1.6\; {\rm syst.}) \times
10^{-40} {\rm cm}^2$ \cite{alb94}.

Another application of the present work is the prediction of the
proton--to--neutron quasi--elastic yield $R(p/n) \equiv \sigma(\nu, \nu'
p)/\sigma(\nu, \nu' n)$ which is currently being measured at the LSND
$^{12}C(\nu, \nu' N)$ experiment. In this experiment $R(p/n)$ is
integrated over $T_F$ and the results are plotted as functions of $G_A^s$
for several values of $F_2^s$. This has been done in Figures 2a and 2b
where results for both neutrino and anti--neutrino scattering are shown
assuming $k_F$ = 225 MeV. It is important to note that for kinematical
reasons the LSND experiment limits the range of integration between $50
\leq T_F \leq 120$ MeV \cite{gar92}. Because of this cut--off imposed by
the experiment, all modifications due to exchange currents for $T_F \leq
50$ MeV are ignored and as a result there is only about 5\% change from
the impulse approximation results in the ratio for the neutrino
scattering  while this change is about 15\% for anti--neutrinos.

To conclude, relativistic exchange current corrections
have been applied to neutrino--nucleus scattering assuming finite
strange quark form factors of the nucleon. The generalized version of the
method of Chemtob and Rho used in this work is so far the
most economical way to estimate exchange current
corrections to low and intermediate energy neutrino--nucleus scattering since
it treats
all the $SU(3)$ vector and axial currents on the same footing.
As examples, soft--pion exchange current corrections have been applied to
quasi--elastic
neutrino--nucleus scattering using a simple Fermi Gas model and
kinematics of the on--going LSND experiment. The differential cross sections
for the
$^{12}C(\nu, \nu' p)$ reaction is found to be sensitive to the values of
the strange quark form factors while the exchange current corrections to
the cross section were found to become more important with increasing
nuclear density. However, because of an experimental kinematical cut,
these exchange current effects are considerably reduced when evaluating
the integrated ratio of proton--to--neutron yields currently being measured at
LSND.
For the charged current case exchange current effects reduce the impulse
approximation results by 5 to 10\% depending on the nuclear density.
Nevertheless, the discrepancy between theory and experiment for the recently
reported $^{12}C(\nu_{\mu^-},\mu^-)X$ total cross section remains
unexplained.
Finally, an extension of the present application to finite nuclei is in
progress.

\acknowledgements

YU and PJM are supported by the foundation for Fundamental Research of
Matter (FOM) and the National Organization for Scientific Research (NWO).
JMU is carrying out the work as a part of a Community
training project financed by the European Commission under Contract
ERBCHBICT 920185.

%
% BIBLIOGRAPHY

\vfill\eject
%
% FIGURE CAPTIONS
%
\centerline{\bf FIGURE CAPTIONS}
\vskip 1cm
FIG~1. \hspace{0.1 cm} a) $^{12}C(\nu,\nu'p)$ differential cross section versus
the kinetic energy of the ejected nucleon, $T_F$ for several values of
Fermi momentum $k_F$. The incident neutrino energy is 200 MeV and the
values for the strange quark form factors are $F_2^s = -0.21$ and $G_A^s
= -0.19$. The long dashed curve is the impulse approximation result while the
solid curves have been obtained with the full exchange current corrections.
Results may be identified by their values at the quasi--elastic peak
around $T_F$ = 30 MeV. For $k_F$ = 200 MeV, there is almost no difference
between the two results and the values at the quasi--elastic peak are
both about $70 \times 10^{-42} {\rm cm^2/MeV}$. At the quasi--elastic peak
the impulse approximation result for $k_F$ = 300 MeV has the value of $60
\times 10^{-42} {\rm cm^2/MeV}$ while the corresponding value with
exchange current corrections is $50 \times 10^{-42} {\rm cm^2/MeV}$.
b) $^{12}C(\nu_{\mu^-},\mu^- p)X$ total cross section obtained by folding
the LSND neutrino flux \cite{alb94} versus $k_F$. The long dashed curve
is the impulse approximation result while the solid curve is obtained
with the full exchange current corrections.
\vskip 0.75cm
FIG.~2 \hspace{0.1 cm} a) Ratios of integrated proton--to--neutron
quasi--elastic yield
for the $^{12}C(\nu,\nu'N)$ reaction as functions of $G_A^s$ for two values
of strange magnetic form factor $F_2^s$.
In each case, the dashed line is the impulse approximation result while
the solid line has been corrected for meson exchange currents. The incident
neutrino
energy is assumed to be 200 MeV for both cases and the range of integration was
chosen to be
$50 \leq T_F \leq 120$ MeV to simulate the LSND experiment \cite{gar92}.
b) Same as in a) but for anti--neutrino scattering.
\end{document}